\documentstyle[12pt]{article}
\newcommand{\bce}{\begin{center}}
\newcommand{\ece}{\end{center}}
\newcommand{\beq}{\begin{equation}}
\newcommand{\eeq}{\end{equation}}
\newcommand{\bea}{\vspace{0.25cm}\begin{eqnarray}}
\newcommand{\eea}{\end{eqnarray}}

\newcommand{\ba}{\begin{array}}
\newcommand{\ea}{\end{array}}

\newcommand{\doublespace}{
    \renewcommand{\baselinestretch}{1.6}\large\normalsize}

\def\lsim{\mathrel{\rlap{\lower4pt\hbox{\hskip1pt$\sim$}}
    \raise1pt\hbox{$<$}}}     %less than or approx. symbol
\def\gsim{\mathrel{\rlap{\lower4pt\hbox{\hskip1pt$\sim$}}
    \raise1pt\hbox{$>$}}}     %greater than or approx. symbol

\def\lsim{\mathrel{\rlap{\lower4pt\hbox{\hskip1pt$\sim$}}
    \raise1pt\hbox{$<$}}}         %less than or approx. symbol
\def\gsim{\mathrel{\rlap{\lower4pt\hbox{\hskip1pt$\sim$}}
    \raise1pt\hbox{$>$}}}         %greater than or approx. symbol

\def\beq{\begin{equation}}
\def\endeq{\end{equation}}
\def\arr{\begin{eqnarray}}
\def\endarr{\end{eqnarray}}

\makeindex
\begin{document}
%\large
\phantom{.}\hspace{8.5cm}{KFA-IKP(TH)-1994-35} \\
\phantom{.}\hspace{8.0cm}{October 1994}
\vspace{2cm}
\begin{center}
{\bf \huge The apparent Coulomb reacceleration of neutrons
in electrodissociation of the deuteron \\}
\vspace{1cm}
{\bf A.Bianconi$^{1,2)}$, S.Jeschonnek$^{3)}$,
N.N.Nikolaev$^{3,4)}$, B.G.Zakharov$^{3,4}$ } \medskip\\
{\small \sl
$^{1)}$Istituto Nazionale di Fisica Nucleare,
Sezione di Pavia, Pavia, Italy \\
$^{2)}$Dipartimento Fisica Nucleare e Teorica,
Universit\`a di Pavia, Italy \\
$^{3)}$IKP(Theorie), Forschungszentrum  J\"ulich GmbH.,\\
D-52425 J\"ulich, Germany \\
$^{4)}$L.D.Landau Institute for Theoretical Physics, \\
GSP-1, 117940, ul.Kosygina 2, V-334 Moscow, Russia
\vspace{1cm}\\}
{\bf \LARGE A b s t r a c t \bigskip\\}
\end{center}
We demonstrate that the final state $p$-$n$ interaction in the
reaction of electrodissociation of the deuteron at large $Q^{2}$
in a static external field leads to the apparent reacceleration
of neutrons. The shift of the neutron velocity from the velocity
of the deuteron beam is related to the quantum-mechanical
forward-backward asymmetry of the missing momentum distribution
in the $^2H(e,e'p)n$ scattering.


%--------------------------------------------------
\newpage
\doublespace

In the recent experiment [1] an intriguing velocity shift between
the neutrons and $^{9}Li$ fragments in the Coulomb dissociation of
$^{11}Li$ nuclei was observed. The observation was qualitatively
interpreted as an effect of classical deceleration of $^{11}Li$
and acceleration of $^{9}Li$ in the Coulomb field and led to
the discussion of the Coulomb reacceleration as a clock for nuclear
reactions [1,2]. The subsequent quantum-mechanical treatment of the
reacceleration effect did not support this classical interpretation,
though ([2,3], see also [4-6]).

In this paper we address the contribution of final state interaction
(FSI) between the fragments of electrodissociation into the shift
of the average velocity of fragments from the beam velocity. We
consider the shift of the velocity of neutrons in the
 simplest case of the Coulomb
dissociation of deuterons at large momentum transfer squared $Q^{2}$.
The deuteron is particularly suited for our investigation as
it is the simplest example for the mechanism of reacceleration by
FSI and as there is a realistic wave function calculated from
the Bonn potential [7] available and FSI in $^2H(e,e'p)n$ is well
understood [8].
The neutron does not
interact with the Coulomb field, and the shift $\Delta v_{z}$ of
the average velocity of neutrons from the velocity of the deuteron
beam only can come from FSI between
the neutron and the proton.
We consider the case of weak Coulomb
field which is treated to the first order, and relate the velocity
shift $\Delta v_{z}$ to the quantum-mechanical FSI effect [8,9] of
the forward-backward asymmetry of the missing momentum distribution
$W(p_{m})$ in the related $^2H(e,e'p)n$ reaction. For the large
$Q^{2}$, we derive a simple relationship between the velocity
shift $\Delta v_{z}$ and the neutron-proton forward scattering
amplitude. Depending on the real part of the $n$-$p$ scattering
amplitude, both the apparent deceleration and acceleration of
neutrons are possible.

The kinematics of the $^2H(e,e'p)n$ reaction is usually described
in the laboratory frame in which the deuteron is at rest, the
virtual photon has the 4-momentum $q=(\nu,\vec{q}\,)$ and $Q^{2}=
-q^{2}$, the struck proton is detected with the momentum $\vec{p}$
and the spectator neutron carries the missing momentum $\vec{p}_{m}
=\vec{q}-\vec{p}$. The same process can also be viewed in the Breit
frame in which $\nu=0$ and $q=(0,0,0,-\sqrt{Q^{2}})$, often used
in the parton model description of deep inelastic lepton-nucleon
scattering. In the Breit frame, the beam of deuterons with velocity
$\vec{v}_{D}$ along the $z$-axis dissociates in a static external
Coulomb field.
In the laboratory frame in the nonrelativistic case,
$\vec v_D = 0$ and
$
\Delta \vec{v}\, ^{lab} = \vec{v}_{n}-\vec{v}_{D} = \vec v_n
= \frac{\vec p_m}{m_n}
\,.$
The average velocity shift $\Delta v_{z}$ equals in the
nonrelativistic case in the Breit frame
\beq
\Delta v_z =
\langle\Delta v_{z}^{Breit}\rangle =
- {1 \over m_{n}} \langle p_{m,z}^{lab} \rangle
\label{eq:1}
\endeq
and vanishes unless the missing momentum distribution $W(\vec{p}_{m})$
has a forward-backward asymmetry.

In the relativistic case, the velocity shift depends on the
reference frame, and it is more convenient to consider the
shift of the average rapidity of neutrons
$y={1\over 2}\log{1+v_{z} \over 1-v_{z}}$ from the rapidity $y_{D}$
of deuterons, $\Delta y = \langle y_{n} \rangle -y_{D} $,
which is the same in all reference frames. As $v_D = 0$ in the
laboratory frame, $y^{lab}_D = 0,$ and
\beq
\Delta y =
- \langle y_n^{lab} \rangle = - \frac {1}{2} \left < \log
\frac{m_n + p_{m,z}^{lab}} {m_n -  p_{m,z}^{lab}} \right >
\approx - \frac {1} {m_n} \langle p_{m,z}^{lab} \rangle
= \Delta v_z
\label{eq:2}
\,,
\endeq
where we have assumed $E^{lab}_n \approx m_n$.

Now we focus on the calculation of the average missing momentum
$\vec{p}_{m}$ in the laboratory frame, where the deuteron is initially
at rest. For the sake of simplicity, we neglect the magnetic interaction
of nucleons and the spin dependence of the proton-neutron scattering
amplitude. Then, for the unpolarized deuterons, one finds the missing
momentum distribution [8]
\arr
W(\vec p_m) =
{1 \over 4\pi(2\pi)^{3}}
\int d^{3}\vec{r} d^{3}\vec{r}\,'\exp[i\vec p_m \cdot
(\vec{r}\,'-\vec r\,)]
S(\vec r\,) S^{\dagger}(\vec r\,')  \nonumber\\
\Bigg[ {u(r) \over r} {u(r') \over r'}\ +\
{1 \over 2} {w(r) \over r} {w(r') \over r'}
\Bigg(3{(\vec r\cdot \vec r\,')^2 \over (rr')^2} - 1\Bigg)\Bigg],
\label{eq:3}
\endarr
where $u/r$ and $w/r$ are the radial wave functions of the S and D
wave states of the deuteron, with the conventional normalization
$\int dr (u^2+w^2) = 1$, and $S(\vec{r}\,)$ is the FSI operator. In
this paper we consider the case of large $Q^{2}$ and high kinetic
energy of the struck proton $T_{kin}\approx Q^{2}/2m_{p}$, such that
FSI can be described by the Glauber theory [10]. Defining the
transverse and longitudinal components $\vec{r}$
$\equiv$ $(\vec{b}+z\hat q)$, where $\vec b$ and $\vec q$ are
orthogonal, we can write
\beq
S(\vec{r}\,) = 1-\theta(z)\Gamma(\vec{b}),
\label{eq:4}
\endeq
where $\Gamma(\vec{b})$ is the profile function of the
proton-neutron scattering. The Glauber theory
representation (\ref{eq:3})
assumes predominantly forward scattering [10], and in the
nucleon-nucleon and nucleon-nucleus scattering it was shown
to hold at $T_{kin} \gsim $ 0.5\,GeV, {\sl i.e.}, at $Q^{2}\gsim
1$GeV$^{2}$ (for a comprehensive review see [11]). At high
energy, the profile function can conveniently be parameterized
as
\beq
\Gamma(\vec{b}) \ \equiv\
{ \sigma_{tot} (1 - i \rho) \over 4 \pi b_{o}^2  }
\exp \Big[-{\vec{b}^2 \over 2 b_{o}^2} \Big] =(1-i\rho)\Gamma_{0}(b)
\, ,
\label{eq:5}
\endeq
where $\rho$ is the ratio of the real to imaginary part of the
forward elastic scattering amplitude, and $b_{o}^{2}$ is the
diffraction slope of elastic scattering, $d\sigma_{el}/dt \propto
\exp(-b_{o}^{2}|t|)$, where $t$ is the square of the momentum
transfer.

Introducing new variables $\vec{s}=\vec{r}\,'-\vec{r}$ and
$\vec{R}={1\over 2}(\vec{r}\,'+\vec{r})$ and integrating by parts,
we can write
\arr
\int d^{3}\vec{p}_{m}\,\vec{p}_{m}W(\vec{p}_{m}) =
{}~~~~~~~~~~~~~~~~~~~~~~~~~~~~~~~\nonumber\\
{i \over 4\pi}
\int d^{3}\vec{R}
{\partial \over \partial \vec{s}}
\left\{
S(\vec{r}\,)
S^{\dagger}(\vec{r}\,')
\left[ {u(r) \over r} {u(r') \over r'}\ +\
{1 \over 2} {w(r) \over r} {w(r') \over r'}
\left(3{(\vec r\cdot \vec r\,')^2 \over (rr')^2} - 1\right)\right]
\right\}
_{\vec{s}=0} \nonumber \\
=\rho \hat q \int d^{2}\vec{b}\,\Gamma_{0}(b)
{1\over 4\pi b^{2}} [ u(b)^{2} + w(b)^{2}] ~~~~~~~~~~~~~~~~~\, ,
\label{eq:6}
\endarr
where $\hat q $ is the unit vector in z direction.
Here we made an explicit use of equation (3) for the FSI operator.
Because of the attenuation of the flux of protons by FSI,
$\int d^{3}\vec{p}_{m}W(\vec{p}_{m}) <1$, but this departure from
unity is small, $\sim 7\%$ [8,12], and to this accuracy
\beq
\langle \vec{p}_{m} \rangle =
{\int d^{3}\vec{p}_{m}\,\vec{p}_{m}W(\vec p_m)\over
\int d^{3}\vec{p}_{m}W(\vec p_m)} \approx
\int d^{3}\vec{p}_{m}\,\vec{p}_{m}W(\vec p_m)  \, .
\label{eq:7}
\endeq
At $T_{kin} \gsim 0.5$\,GeV, the radius of $n$-$p$ FSI is small.
Typically, $b_{o} \sim 0.5$\,fm [11,13], and we have a strong
inequality $b_{o}^{2} \ll R_{D}^{2}$, where $R_{D}\sim 2$\,fm is
the deuteron radius. This leads to a simple estimate
\beq
\Delta v_{z}
 \sim - \rho{\sigma_{tot}(np) \over
 4 \pi m_{n}R_{D}^{3}} \, ,
\label{eq:8}
\endeq
which is not sensitive to $b_{o}$.

This apparent reacceleration of
neutrons is the purely quantum-mechanical effect of an interference
between the plane wave and FSI components of the electrodissociation
amplitude [8].
We emphasize that it is not the effect of higher orders in an
external Coulomb field.
The sign of the velocity shift only depends on the sign
of the real part of the forward proton-neutron scattering amplitude.
The Glauber approximation considered here is applicable when many
partial waves contribute to the $n-p$ FSI. The opposite limiting case
of small $Q^{2}$ and very low excitation energies, when FSI effects
can be modelled by the zero-range interaction which takes place only
in the $S$-wave, was considered in [3]. In Ref.~[3] a similar
conclusion that the sign of the reacceleration effect depends on
the sign of the scattering phase, was reached. With $\sigma_{tot}(np)
\sim 40$\,mb, $\rho \sim -0.4$ and $b_{o}\sim 0.5$\,fm, typical of
the $N$-$N$ interaction at $T_{kin} \gsim 1.0$\,GeV [10,12,13],
equation (\ref{eq:7}) gives the shift of the velocity
of neutrons $\Delta v_{z} \sim  0.0026 $, which is very close
to the estimate (8). The estimate (8) makes it obvious that
$\Delta v_z$ is very small because of the deuteron being a
dilute system, which makes the $n-p$ FSI weak.
 The results of numerical
calculations with the realistic Bonn wave function [7] are
presented in Fig.~1. The parameters $ \sigma_{tot}, \rho$ and
$b_o$ have been taken from [13,14].
The $Q^2$ dependence of $\langle \Delta v_z \rangle$
is due to the energy dependence of $\rho$ as given by
dispersion relation calculations as reported in [14].

To summarize, the purpose of this note has been a derivation
of the FSI generated effect of the apparent reacceleration of
neutrons in the electrodissociation of fast deuterons in a static
external field. We related the apparent reacceleration to the
quantum-mechanical effect of the FSI induced forward-backward
asymmetry of the missing momentum distribution in the related
$^2H(e,e'p)n$ scattering.
The quantal physics of the apparent reacceleration
by FSI is simple and both the acceleration and deceleration
effects are possible depending on the sign of the real part of
the $p$-$n$ forward scattering amplitude.
The forward-backward asymmetry of $W(\vec p_m)$ and
$\langle \vec p_{m,z} \rangle$ can, of course, be directly measured
in the traditional $^2H(e,e'p)n$ experiments.
The measurement of the velocity shift
$\langle \Delta v_z \rangle$ in the Coulomb dissociation of
reltivistic deuterons is not feasible, at least at the large $Q^2$
of interest for the present formalism. The point is that for the
finite, and large, size of nuclei the electromagnetic dissociation
of deuterons will be masked by diffraction dissociation
$ dA \rightarrow (pn)A, (pn)A^* $, induced by strong nuclear
interaction [6,15].

{\bf Acknowledgments:}
This work has been done during a
visit of A.B. at IKP, KFA J\"ulich (Germany), supported by
IKP and by INFN (Italy), and in part
by the Vigoni Program of DAAD (Germany) and of the Conferenza
Permanente dei Rettori (Italy).
A.B. thanks J.Speth for the hospitality
at IKP.
This work was also supported by
the INTAS Grant No. 93-239.
\pagebreak\\
% = = = = = = = = = = = = = = = = = = = = = = = = = = = = = = = = = = =

\pagebreak

{\bf \Large Figure captions:}

\begin{itemize}

\item[{\bf Fig.~1}]
   The shift of the average velocity of neutrons $\Delta v_{z}$
from the velocity of deuterons
in the electrodisintegration of deuterons {\sl vs.} $Q^{2}$.

\end{itemize}

\begin{thebibliography}{299}
%wwwwwwwww

\bibitem{1}
K.Ieki et al., {\sl Phys. Rev. Lett.} {\bf 70} (1993) 730;
D.Sackett et al., {\sl Phys. Rev.} {\bf C48} (1993) 118.

\bibitem{2}
G.F.Bertsch and C.A.Bertulani, {\sl Nucl. Phys. } {\bf A556} (1993)
136, C.A.Bertulani and G.F.Bertsch, {\sl Phys. Rev.} {\bf C49}
(1994) 2389.

\bibitem{3}
S.Typel and G.Baur, {\sl J\"ulich preprint} {\bf KFA-IKP-1993-28}.

\bibitem{4}
R.Shyam, P.Banerjee and G.Baur, {\sl Nucl. Phys.} {\bf A540} (1992)
341.

\bibitem{5}
G.Baur, C.A.Bertulani and D.M.Kalassa, {\sl Nucl. Phys.} {\bf A550}
(1992) 527.

\bibitem{6}
P.Banerjee and R.Shyam, {\sl Nucl. Phys.} {\bf A561} (1993) 112.

\bibitem{7}
R.Machleidt, K.Holinde, and C.Elster, {\sl Phys. Rep.}
{\bf 149} (1987), 1.

\bibitem{8}
A.Bianconi, S.Jeschonnek, N.N.Nikolaev, and B.G.Zakharov,
J\"ulich preprint {\bf KFA-IKP(Th)-1994-34} (1994),
submitted to {\sl Phys. Lett.} {\bf B}.

\bibitem{9}
A.Bianconi, S.Jeschonnek, N.N.Nikolaev, and B.G.Zakharov,
J\"ulich preprint {\bf KFA-IKP(Th)-1994-29} (1994),
{\sl Phys. Lett.} {\bf B}, in press.

\bibitem{10}
R.J.Glauber, in: {\sl Lectures in Theoretical Physics}, v.1,
ed. W.Brittain and L.G.Dunham. Interscience Publ., N.Y., 1959;
R.J.Glauber and G.Matthiae, {\sl Nucl. Phys.} {\bf B21} (1970) 135.

\bibitem{11}
G.D.Alkhazov, S.I.Belostotsky, and A.A.Vorobyev, {\sl Phys. Rep.}
{\bf C42} (1978) 89.

\bibitem{12}
N.N.Nikolaev, A.Szczurek, J.Speth, J.Wambach, B.G.Zakharov, and
V.R.Zoller, {\sl Phys. Rev.} {\bf C50} (1994) R1296.

\bibitem{13}
T.Lasinski et al., {\sl Nucl. Phys.} {\bf B37} (1972) 1.

\bibitem{14}
C.Lechanoine-LeLuc and F.Lehar, {\sl Rev. Mod. Phys.} {\bf 65}, 47
(1993).

\bibitem{15}
A.I.Akhiezer and A.G.Sitenko, {\sl Phys. Rev.} {\bf 106} (1957) 1236.

\end{thebibliography}
\end{document}